\documentclass[aps,twocolumn]{revtex4-2}
\pdfoutput=1
\usepackage{amsmath}
\usepackage{amssymb}
\usepackage{amsfonts}
\usepackage{graphicx}
\usepackage{booktabs}
\usepackage{hyperref}

\DeclareMathOperator*{\argmin}{arg\,min}

\newcounter{theorem}

\begin{document}

\title{Logical-Space Theory of Ordered-Statistics Decoding for Quantum LDPC Codes}

\author{Yun Shang}
\email{shangyun@amss.ac.cn}
\affiliation{Institute of Mathematics, Academy of Mathematics and Systems Science,
Chinese Academy of Sciences, Beijing 100190, China}
\affiliation{State Key Laboratory of Mathematical Sciences, Academy of Mathematics
and Systems Science, Chinese Academy of Sciences, Beijing 100190, China}

\begin{abstract}
Quantum decoding succeeds when a correction matches the logical class of the
error, whereas ordered-statistics decoding (OSD) searches a finite list of
physical corrections.  We connect these spaces through a rank-$k$
logical-signature map $M_{\rm log}:\mathbb F_2^q\to\mathbb F_2^k$.  For fixed,
error-independent orderings it yields the exact logical-signature distribution,
the OSD-0 logical error rate, and the logical-class coverage of any fixed
list.  For minimum-cost list decoding we prove
$P_L=P_{\rm miss}+P_{\rm sel}$, separating absent-class from wrong-selection
failures.  Hypergraph-product and bivariate-bicycle codes are consistent
with the coverage predictions.  On the $[\![144,12,12]\!]$ gross code, 672
coverage-aware candidates outperform 899 combination-sweep candidates (paired McNemar
$p=0.013$), while two lists with identical logical coverage give
$P_L=11.30\%$ and $3.78\%$; the signature-based coverage compression
persists across 30 random column orderings.  Logical coverage and
within-class representative selection are therefore complementary resources
that candidate count alone does not capture. The framework guides OSD list construction and compression by identifying coverage requirements and distinguishing coverage losses from failures of representative selection.
\end{abstract}

\maketitle

\section{Introduction}

The objective of a quantum decoder is intrinsically different from that of a
classical decoder.  It need not identify the physical error itself: any
correction in the same logical equivalence class succeeds.  This degeneracy is
central to stabilizer and CSS quantum error correction~\cite{CalderbankShor1996,
Steane1996,IyerPoulin2015}, yet many practical decoders generate and rank
candidates in the much larger space of physical errors.  Understanding how a
finite physical candidate list resolves logical equivalence classes is
therefore a basic question for decoder design.  This question matters for qLDPC codes, whose sparse constructions range from early
sparse-graph and hypergraph-product families~\cite{MacKay2004,TillichZemor2014}
to asymptotically good constructions and the high-rate finite codes used in
fault-tolerance proposals~\cite{PanteleevKalachev2022,Bravyi2024}.

Ordered-statistics decoding (OSD), originally developed as a soft-decision
method for classical linear codes~\cite{FossorierLin1995}, makes this mismatch
particularly explicit.  Belief propagation (BP) is attractive for sparse
quantum codes but is challenged by short cycles and quantum
degeneracy~\cite{PoulinChung2008}.  BP combined with OSD has consequently become
a widely used qLDPC benchmark and post-processing strategy
~\cite{Roffe2020,Panteleev2021,Valls2021}.  OSD uses Gaussian elimination
to parameterize syndrome-compatible corrections by free variables, evaluates
a prescribed finite candidate list, and returns its minimum-cost member.  Its
performance is usually characterized by Monte Carlo logical error rates or by
the size and order of the physical candidate list.  What is missing is an
algebraic description of which \emph{logical} classes such a list can reach and
how that structure predicts failure.

We establish three results.  \emph{(i) Representation and prediction.}  A
surjective logical-signature map
$M_{\rm log}:\mathbb F_2^q\to\mathbb F_2^k$ from OSD free variables to
logical space reduces success to $M_{\rm log}f=M_{\rm log}e_F$; for fixed,
error-independent column orderings its spectrum gives the exact
logical-signature distribution, the OSD-0 logical error rate, and the
coverage probability of any fixed candidate list.  \emph{(ii) Design.}  We
organize candidate lists by their logical signatures, deriving a coverage
lower bound and an explicit greedy construction that compresses the
required weight-2 prefix, and we verify the compression across 30 random
column orderings.  \emph{(iii) Mechanism.}  A class-energy construction
yields the exact decomposition $P_L=P_{\rm miss}+P_{\rm sel}$ and a
syndrome-level decision margin; controlled comparisons between lists with
identical coverage, together with a per-shot analysis, show how
within-class representatives determine the selection term.

The exact distributional results hold for fixed, error-independent column
orderings; Sec.~\ref{sec:discussion} records what remains valid under
syndrome- or BP-dependent orderings.

Related qLDPC decoders include small-set flipping, hybrid BP, and beam
search~\cite{LeverrierTillichZemor2015,Grospellier2021,BeamSearch2025}.
Other approaches explicitly aggregate or approximate logical-coset
information: the Frontier decoder of Leverrier and
Urbanke~\cite{Frontier2026} approximates logical-coset posterior masses by
dynamic programming over residual-syndrome prefixes, while the coset
ensemble decoder of Liang
\emph{et al.}~\cite{COSSETTE2026} aggregates candidates across randomized
Union-Find forests.  Our aim is different: we expose the algebraic logical-space
structure of the OSD free-variable representation itself, yielding exact
coverage predictions and structural diagnostics for finite OSD candidate
sets.

The rest of the paper is organized as follows.  Sec.~\ref{sec:rep} sets up
the logical-coordinate representation of OSD and proves its main
properties.  Sec.~\ref{sec:dist} gives the exact logical-signature
distribution, the OSD-0 error rate, and the coverage probabilities of a
fixed list.  Sec.~\ref{sec:covsel} decomposes logical failure into
coverage and representative selection.  Sec.~\ref{sec:list} turns
coverage into a list-construction problem, giving the coverage lower
bound, the greedy construction, and the coverage-minimal list.
Sec.~\ref{sec:num} presents the numerical results, and
Sec.~\ref{sec:discussion} discusses scope, computational cost, and open
problems.

\section{Logical-Coordinate Representation of OSD}
\label{sec:rep}

\subsection{Code Parameters and RREF Decomposition}

We consider $X$-error decoding for a CSS quantum code.
Let $H_Z \in \mathbb{F}_2^{m_Z \times n}$ be the $Z$-parity-check matrix,
$H_X \in \mathbb{F}_2^{m_X \times n}$ the $X$-parity-check matrix,
and $L_Z \in \mathbb{F}_2^{k \times n}$ a basis of logical $Z$ operators.
The code encodes $k$ logical qubits into $n$ physical qubits.
For $X$-error decoding, syndrome-zero errors form $\ker H_Z$ and logically
trivial syndrome-zero errors form $\operatorname{row}H_X$.  We choose $L_Z$
so that
$\ker(L_Z|_{\ker(H_Z)}) = \operatorname{row}(H_X)$,
i.e., $L_Z$ separates the logical quotient $\ker(H_Z) / \operatorname{row}(H_X) \cong \mathbb{F}_2^k$.

Given a fixed, error-independent column ordering (e.g., sorted by a priori reliability),
Gaussian elimination on $H_Z$ yields the reduced row-echelon form.
Let $r = \operatorname{rank}(H_Z)$ and $q = n - r$.
After removing redundant (all-zero) rows, the nonzero part of the RREF takes the form
\begin{equation}
[\,I_r \mid A\,], \qquad A \in \mathbb{F}_2^{r \times q}.
\end{equation}
The $q$ nullspace vectors
$g_j = [A e_j;\, e_j] \in \mathbb{F}_2^n$
associated with the free coordinates form a basis of $\ker(H_Z)$.

We denote by $G_F \in \mathbb{F}_2^{n \times q}$ the matrix whose columns are
the physical error patterns $g_j$ (in the original column ordering).
For any free-variable pattern $f \in \mathbb{F}_2^{q}$, $G_F f \in \ker(H_Z)$
is the corresponding physical error.

\subsection{The Logical-Signature Map}

\begin{figure*}[t]
\centering
\includegraphics[width=0.92\textwidth]
{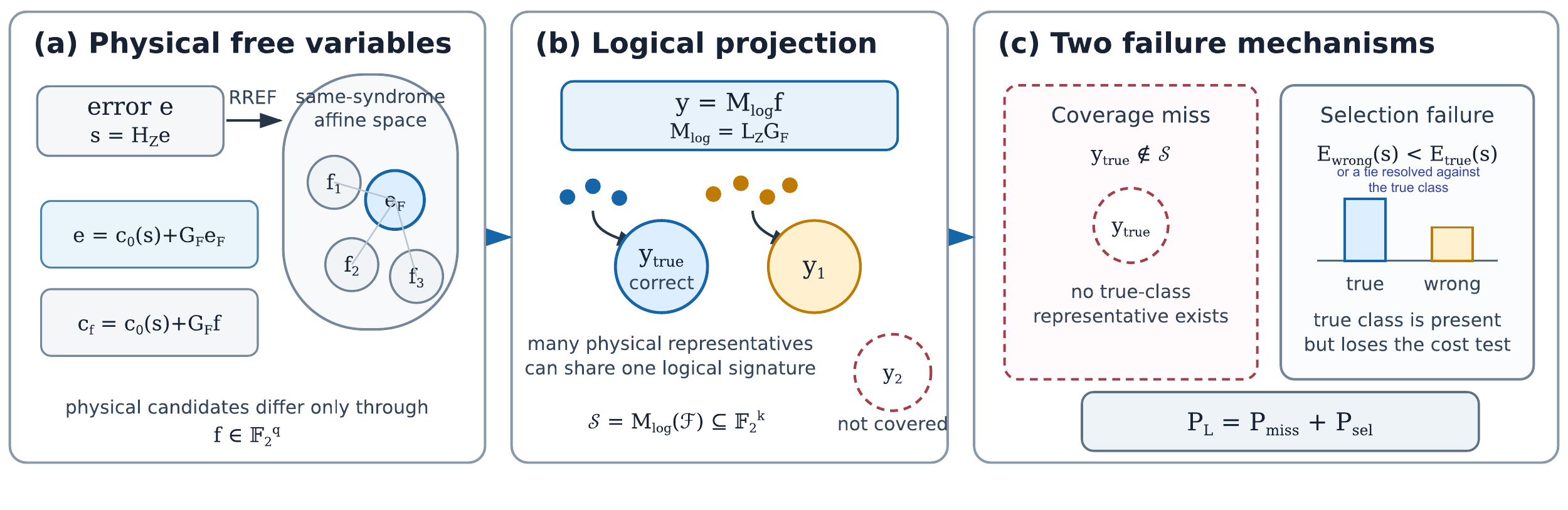}
\caption{
Logical-space structure of finite-list OSD.
(a) Gaussian elimination parametrizes syndrome-compatible corrections
as $c_f=c_0(s)+G_Ff$.
(b) $M_{\rm log}=L_ZG_F$ maps free-variable patterns onto logical
displacements (many-to-one).
(c) Logical failure: coverage miss ($y_{\rm true}\notin\mathcal S$) or
selection failure (a competing class has lower cost, or equal cost but is
selected by the tie rule).
Hence $P_L=P_{\rm miss}+P_{\rm sel}$.
}
\label{fig:theory}
\end{figure*}

\begin{quotation}
\noindent\refstepcounter{theorem}\textbf{Theorem \thetheorem\ (Logical-coordinate representation).}
\label{thm:1}
Let $M_{\rm log} = L_Z G_F \in \mathbb{F}_2^{k \times q}$
be the logical-signature matrix, where $G_F$ is a basis of $\ker H_Z$
from RREF elimination and $L_Z$ is a basis of logical $Z$ operators.
For any error $e \in \mathbb{F}_2^n$ with syndrome $s = H_Z e$,
decompose $e = c_0(s) + G_F e_F$ where $c_0(s)$ solves $H_Z c_0(s) = s$
and $e_F \in \mathbb{F}_2^{q}$.
For any candidate $f \in \mathcal{F} \subset \mathbb{F}_2^{q}$,
let $c_f = c_0(s) + G_F f$.
Then
\begin{equation}
e + c_f = G_F(e_F + f),
\label{eq:residual}
\end{equation}
and the logical outcome is
\begin{equation}
\begin{aligned}
\text{logical success}
&\Longleftrightarrow M_{\rm log}(e_F + f) = 0 \\
&\Longleftrightarrow M_{\rm log} f = M_{\rm log} e_F.
\end{aligned}
\label{eq:success}
\end{equation}
Moreover, $\operatorname{rank}(M_{\rm log}) = k$.
\end{quotation}

\textit{Proof.}
Eq.~\eqref{eq:residual} follows directly from $e=c_0(s)+G_Fe_F$ and
$c_f=c_0(s)+G_Ff$.  Since $G_F$ is an isomorphism from $\mathbb F_2^q$
onto $\ker H_Z$, the image of $M_{\rm log}=L_Z\circ G_F$ equals the image
of $L_Z|_{\ker H_Z}$.  By the first isomorphism theorem and
$\ker(L_Z|_{\ker H_Z})=\operatorname{row}H_X$,
$\operatorname{im}M_{\rm log}\cong\ker H_Z/\operatorname{row}H_X\cong
\mathbb F_2^k$, hence $\operatorname{rank}M_{\rm log}=k$.

The columns of $M_{\rm log}$ classify free variables into
\emph{dangerous} ($M_{{\rm log},j} \neq 0$) and \emph{safe} ($M_{{\rm log},j} = 0$)
columns. Flipping a dangerous free variable changes the logical displacement;
flipping a safe one (a stabilizer in $\operatorname{row}(H_X)$) does not.
The number of dangerous columns is $n_{\rm dangerous} = |\{j : M_{{\rm log},j} \neq 0\}|$.

Theorem~\ref{thm:1} reduces decoding to signature matching:
the list contains a successful candidate---a pattern $f$ with the same
logical displacement as the true error---exactly when
$M_{\rm log}e_F \in \mathcal{S}=M_{\rm log}(\mathcal{F})$.
Whether decoding actually succeeds further depends on the cost rule
selecting such a candidate, a distinction made precise in
Sec.~\ref{sec:covsel}.
OSD list design then becomes a geometric problem in $\mathbb{F}_2^{q}$:
cover $M_{\rm log} e_F$ with some $f \in \mathcal{F}$.

\subsection{Combination-Sweep Candidate Sets}

For fixed, error-independent column ordering, we use the combination-sweep
candidate list $\mathcal{F}_{\rm CS}(\lambda)$: the zero pattern, all $q$
single-coordinate patterns, and all weight-2 combinations of the first
$\lambda$ columns (with $\lambda \le q$),
\begin{equation}
\begin{aligned}
\mathcal{F}_{\rm CS}(\lambda) = \{0\} &\cup
\{e_i : 0 \le i < q\} \\
&\cup \{e_i + e_j : 0 \le i < j < \lambda\},
\end{aligned}
\end{equation}
where $e_i \in \mathbb{F}_2^{q}$ is the $i$-th unit vector.
Here $\lambda$ controls the prefix length of the pair sweep, not a
Hamming-weight threshold: $\mathcal{F}_{\rm CS}(\lambda)$ does not enumerate
all patterns of weight at most $\lambda$.
The candidate count is
\begin{equation}
N_{\rm cand}(\lambda) = 1 + q + \binom{\min(\lambda,q)}{2}.
\end{equation}
The OSD-0 family is $\mathcal{F}_0 = \{0\}$.

\section{Exact Logical Distributions and Coverage}
\label{sec:dist}

The baseline decoder OSD-0 selects the correction
obtained by setting all $q$ free coordinates to zero in the Gaussian elimination
($f = 0$).
This succeeds if and only if $M_{\rm log} e_F = 0$, i.e., the true error has
trivial logical displacement in the free-variable decomposition.

\begin{quotation}
\noindent\refstepcounter{theorem}\textbf{Theorem \thetheorem\ (Exact OSD-0 logical error rate).}
\label{thm:2}
Let $c_0(s)$ be the elimination baseline with all free coordinates set to zero,
and let $e = c_0(s) + G_F e_F$ be the corresponding decomposition of the true error.
Assume a \emph{fixed, error-independent} column ordering.
Under an i.i.d.\ bit-flip channel with error rate $p$,
the free-variable component satisfies $(e_F)_j \sim_{\rm iid} \operatorname{Bernoulli}(p)$
for $j = 0, \ldots, q-1$.
Let $Y = M_{\rm log} e_F$ denote the logical displacement and let
$u \in \mathbb{F}_2^k$.
Then
\begin{multline}
\Pr(Y = u) = 2^{-k} \sum_{v \in \mathbb{F}_2^k}
(-1)^{u \cdot v}\\
\times (1-2p)^{|M_{\rm log}^T v|},
\label{eq:fourier}
\end{multline}
where $|x|$ denotes the Hamming weight.
Consequently, the OSD-0 logical error rate is
\begin{equation}
P_L^{\rm OSD-0} = 1 - 2^{-k} \sum_{v \in \mathbb{F}_2^k}
(1-2p)^{|M_{\rm log}^T v|}.
\label{eq:osd0}
\end{equation}
For small $p$,
\begin{equation}
P_L^{\rm OSD-0} = n_{\rm dangerous}\, p + O(p^2).
\label{eq:osd0small}
\end{equation}
\end{quotation}

\textit{Proof.}
Let $Y=M_{\rm log}e_F$ and write $M_j$ for the $j$th column of
$M_{\rm log}$.  For $v\in\mathbb F_2^k$ the characteristic function
factors as
\begin{align}
\mathbb E[(-1)^{v\cdot Y}]
&=\prod_{j=0}^{q-1}\mathbb E[(-1)^{(v\cdot M_j)(e_F)_j}]
\notag\\
&=(1-2p)^{|M_{\rm log}^Tv|},
\label{eq:characteristic}
\end{align}
since the factor for column $j$ is $1$ when $v\cdot M_j=0$ and $1-2p$
when $v\cdot M_j=1$.  Walsh inversion over $\mathbb F_2^k$ gives
Eq.~\eqref{eq:fourier}.  OSD-0 contains only $f=0$ and succeeds iff
$Y=0$, hence Eq.~\eqref{eq:osd0}.  Expanding
$(1-2p)^w=1-2wp+O(p^2)$ and observing that a nonzero column $M_j$
satisfies $v\cdot M_j=1$ for exactly half of all $v$ gives
$\sum_v|M_{\rm log}^Tv|=2^{k-1}n_{\rm dangerous}$ and hence
Eq.~\eqref{eq:osd0small}.

Given a candidate set $\mathcal{F}$, the \emph{covered logical displacements} are
\begin{equation}
\mathcal{S} = M_{\rm log}(\mathcal{F}) = \{M_{\rm log} f : f \in \mathcal{F}\}
\subseteq \mathbb{F}_2^k.
\end{equation}
In the fixed-order iid setting of Theorem~\ref{thm:2}, the analytical coverage is
\begin{equation}
P_{\rm cover} = \sum_{y \in \mathcal{S}} \Pr(M_{\rm log} e_F = y).
\end{equation}

In the same fixed-order iid setting, uncovered contributions are enumerated by
the weight spectrum of $M_{\rm log}$ outside $\mathcal{S}$.
Define, for each weight $w \in \{0, \ldots, q\}$,
\begin{equation}
B_w = |\{x \in \mathbb{F}_2^{q} : |x| = w,\; M_{\rm log} x \notin \mathcal{S}\}|.
\end{equation}
Since the candidate set $\mathcal{F}_{\rm CS}(\lambda)$ includes all weight-0 and
weight-1 patterns by default, $B_0 = B_1 = 0$.
The coverage-miss probability $P_{\rm miss}=\Pr(y_{\rm true}\notin\mathcal S)$
admits the exact binomial expansion
\begin{equation}
P_{\rm miss} = \sum_{w=0}^{q} B_w\, p^{w} (1-p)^{q-w}.
\end{equation}
Let $t_{\rm cov} = \min\{w : B_w > 0\}$ be the lowest uncovered weight.
If $B_w = 0$ for every $w$, we set $t_{\rm cov} = \infty$, in which case
$\mathcal S$ contains every logical displacement and $P_{\rm miss}\equiv 0$;
this is realized by the HGP $[\![13,1,3]\!]$ row of the cross-code table in
the Supplemental Material~\cite{Supplemental}.
For finite $t_{\rm cov}$,
\begin{equation}
P_{\rm miss} = B_{t_{\rm cov}}\, p^{t_{\rm cov}} + O(p^{t_{\rm cov}+1}).
\label{eq:pmissasym}
\end{equation}
For any candidate set containing all weight-0 and weight-1 patterns, $t_{\rm cov} \ge 2$
(including $t_{\rm cov}=\infty$), so the dominant uncovered contribution (if any)
enters at order $p^2$ or higher.

\section{Coverage and Representative Selection}
\label{sec:covsel}

\begin{quotation}
\noindent\refstepcounter{theorem}\textbf{Theorem \thetheorem\ (Coverage--selection decomposition).}
\label{thm:3}
For any candidate-based minimum-cost decoder with candidate set
$\mathcal{F}$ and covered displacements $\mathcal{S}$,
define the true logical displacement $y_{\rm true} = M_{\rm log} e_F$,
the logical-class energy
\begin{equation}
E_y(s) = \min_{f \in \mathcal{F}:\, M_{\rm log} f = y}
C[c_0(s) + G_F f],
\end{equation}
with $E_y(s) = \infty$ if no $f \in \mathcal{F}$ satisfies $M_{\rm log} f = y$,
and the cost function $C[\cdot]$ (e.g., weighted Hamming weight).
Ties in minimum cost are broken by the fixed order of the candidate list
$\mathcal{F}$ (the earliest list member attaining the minimum cost).
Then the logical error rate decomposes exactly as
\begin{equation}
\boxed{P_L = P_{\rm miss} + P_{\rm sel}},
\label{eq:decomp}
\end{equation}
where $\widehat y$ denotes the decoder's selected displacement, the
$\argmin$ of $E_y(s)$ over $y \in \mathcal{S}$, and
\begin{align}
P_{\rm miss} &= \Pr(y_{\rm true} \notin \mathcal{S}), \label{eq:Pmiss} \\
P_{\rm sel} &= \Pr\!\big(y_{\rm true} \in \mathcal{S},\;
\widehat y \ne y_{\rm true}\big). \label{eq:Psel}
\end{align}
$P_{\rm cover} = 1 - P_{\rm miss} = \Pr(y_{\rm true} \in \mathcal{S})$
is the coverage probability.
\end{quotation}

\textit{Proof.}
If $y_{\rm true}\notin\mathcal S$, the true logical class has no
representative in the list, so decoding fails; this is the coverage-miss
event of Eq.~\eqref{eq:Pmiss}.  If $y_{\rm true}\in\mathcal S$, decoding
succeeds exactly when the selected minimum-cost class equals $y_{\rm true}$
(ties resolved by the fixed candidate-list order) and fails otherwise;
this is the selection event of Eq.~\eqref{eq:Psel}.  The two events are
disjoint and exhaust all logical failures, hence Eq.~\eqref{eq:decomp}.

Theorem~\ref{thm:3} makes precise the two distinct ways a finite list can
fail. The list contains a candidate with the true logical displacement if and only if \(y_{\mathrm{true}}\in S\). A selection failure occurs when such a candidate is present but the decoder selects a candidate with a different logical displacement. Coverage and representative selection are therefore separate
design resources, treated separately in Secs.~\ref{sec:list} and
\ref{sec:num}.

Define the exact logical decision margin:
\begin{equation}
\boxed{\Delta(s) = \min_{y \neq y_{\rm true}} E_y(s) - E_{y_{\rm true}}(s)}.
\end{equation}
Thus $\Delta>0$ certifies success, $\Delta<0$ failure, and $\Delta=0$ is decided
by the deterministic tie rule of Theorem~\ref{thm:3} (the earliest list member
attaining the minimum cost).  Because $y_{\rm true}$ is required, $\Delta$ is a
diagnostic,
not an online decoder statistic.

\section{Coverage-Aware List Construction}
\label{sec:list}

For the combination-sweep candidate sets $\mathcal{F}_{\rm CS}(\lambda)$,
the covered signatures depend on the first \(\lambda\) free columns used for weight-2 pairs.
We define the \emph{weight-2 coverage threshold}
\begin{equation}
\lambda_2^* = \min\{\lambda : B_2(\lambda) = 0\},
\end{equation}
where $B_2(\lambda)$ uses $\mathcal{S} = M_{\rm log}(\mathcal{F}_{\rm CS}(\lambda))$.
For the gross code under the canonical (reliability-sorted) ordering,
$\lambda_2^* = 78 = q$: the combination-sweep list must include all free columns
to eliminate weight-2 coverage failures.

For a fixed free-variable decomposition, define the prefix-independent quantity
$\Lambda_2(M_{\rm log})$ as the minimum size of a column subset $T$ such that
all pair displacements not already covered by the baseline and single-column signatures are realized by pairs within $T$:
\begin{equation}
\begin{aligned}
\Lambda_2 = \min\big\{|T| : &\; \{M_i \oplus M_j : i < j\} \\
&\subseteq \mathcal S_0 \cup \{M_a \oplus M_b : a,b \in T\}\big\},
\end{aligned}
\end{equation}
where $M_j$ is the $j$th column of $M_{\rm log}$ and
$\mathcal S_0=\{0\}\cup\{M_j:0\le j<q\}$ contains the baseline and all
single-column signatures.  Write
$\mathcal S_{\rm pair}=\{M_i\oplus M_j:i<j\}$ for the set of all pair
displacements and $R=\mathcal S_{\rm pair}\setminus\mathcal S_0$ for the
set of displacements left uncovered by the baseline and the singles.
The full pair-augmented list (baseline, singles, and all pairs) covers
$\mathcal S_{\rm full}=\mathcal S_0\cup\mathcal S_{\rm pair}$.

The implementation uses a deterministic greedy covering procedure.  It first selects the
column that realizes the largest number of uncovered displacements of $R$
with any other column; thereafter it repeatedly adds the column whose pair
displacements with the already-selected columns cover the largest number of
as-yet-uncovered displacements of $R$.  In both steps the score counts
\emph{column pairs} (each pair contributing its displacement, counted with
multiplicity over pairs, not over distinct signatures): writing $T$ for the
selected columns and $U$ for the as-yet-uncovered displacements, the gain of
a column $c$ is
$g(c)=\bigl|\{t\in T: M_c\oplus M_t\in U\}\bigr|$
in the growth step, with the initial selection maximizing the analogous
count over all columns not in $T$.  If no remaining column adds any
uncovered displacement, a new component is opened by the pair of remaining
columns whose pair displacement lies in the uncovered set and which
maximizes the number of further uncovered displacements reachable with the
selected columns.  Ties are broken by the smallest column index
(lexicographically smallest pair), and the construction stops once every
displacement of $R$ is covered; completeness is verified by re-enumeration.
This reports the feasible size $\Lambda_2^{\rm greedy}$, an upper bound on
the exact minimum.  Each pair inside $T$ can add
at most one new element of $R$, so the necessary counting bound
$\binom{|T|}{2}\ge |R|$
gives $|T|\ge35$ for the canonical gross-code decomposition.  The greedy
construction finds $|T|=41$ there, whereas the combination-sweep prefix
threshold is $\lambda_2^*=78=q$.  Robustness across 30 random column
orderings is reported in Sec.~\ref{sec:robust}.

The same counting logic fixes the minimum size of a coverage-complete
candidate list.

\begin{quotation}
\noindent\refstepcounter{theorem}\textbf{Proposition \thetheorem\ (Coverage-minimal completion).}
\label{prop:covmin}
Fix the zero pattern and all $q$ single-coordinate patterns, whose covered
set is $\mathcal S_0$, and let $\mathcal S_{\rm full}=\mathcal S_0\cup
\mathcal S_{\rm pair}$ be the coverage of the full pair-augmented list.  Each
additional candidate contributes at most one new signature, so at least
$|\mathcal S_{\rm full}\setminus\mathcal S_0|$ additional candidates are
required to cover $\mathcal S_{\rm full}$, and choosing one weight-2 pair
for each missing signature attains this bound.  For the canonical
gross-code decomposition $|\mathcal S_{\rm full}|=635$ and
$|\mathcal S_0|=42$, giving
\begin{equation}
N_{\min}=1+78+593=672.
\end{equation}
This minimality is with respect to the fixed baseline and singles and the
target coverage set $\mathcal S_{\rm full}$; it is not a claim of optimal
decoding or minimal logical error rate.
\end{quotation}

For each missing signature the pair is chosen with minimum
\emph{intrinsic weight} $|G_F(e_i+e_j)|$, the Hamming weight of the pair's
own physical realization, with ties broken by the lexicographic order of
$(i,j)$.  Decoder~F of Sec.~\ref{sec:main} implements this construction.

\section{Numerical Results}
\label{sec:num}

\subsection{Cross-Code Validation}

\begin{figure}[t]
\centering
\includegraphics[width=0.85\columnwidth]{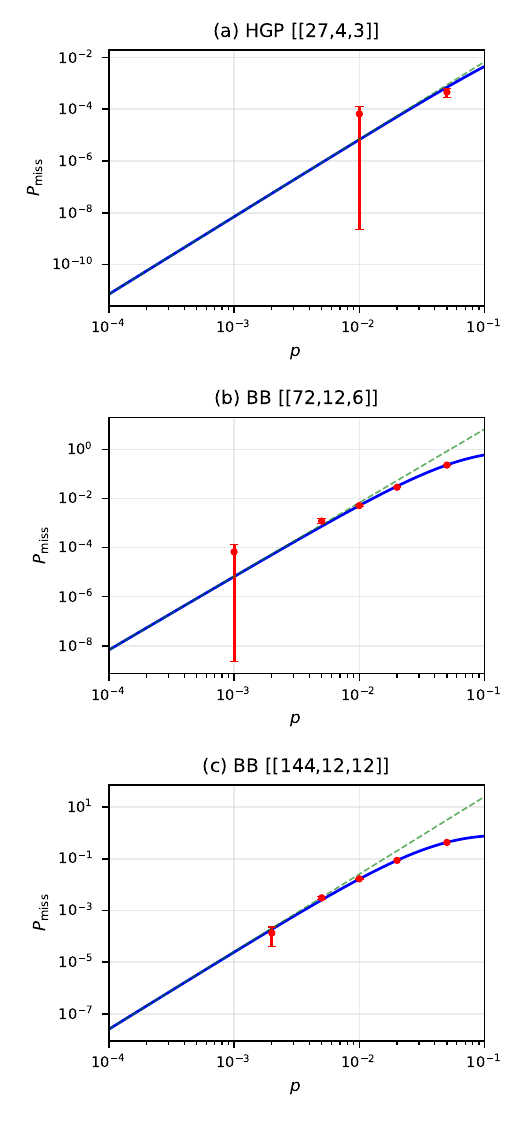}
\caption{$P_{\rm miss}(p)$ for HGP $[\![27,4,3]\!]$, BB $[\![72,12,6]\!]$,
and BB $[\![144,12,12]\!]$.
(HGP $[\![13,1,3]\!]$ omitted: $P_{\rm miss}\equiv 0$ for this code.)
Solid: exact Fourier result; points: Monte Carlo ($1.5\times10^4$ shots/point)
with error bars of $\pm1$ standard error $\sqrt{\hat p(1-\hat p)/N}$ of the
binomial proportion, $N=1.5\times10^4$ (Wald standard error, not a
confidence interval); dashed: $B_3p^3$ small-$p$ asymptote.
Zero-event points are invisible on the logarithmic axis;
few-event bars: see text.}
\label{fig:pmiss}
\end{figure}

We test two HGP and two BB codes under fixed, error-independent orderings;
Fig.~\ref{fig:pmiss} shows $P_{\rm miss}(p)$ for the three codes with nonzero
$P_{\rm miss}$ (HGP $[\![13,1,3]\!]$ has $P_{\rm miss}\equiv 0$).
Table~\ref{tab:cross} collects the structural metrics that drive these
curves: the free dimension $q$, the logical dimension $k$, the number of
dangerous columns, the greedy weight-2 coverage set, and the
covered-displacement count $|\mathcal S_{\rm full}|$ of the full
pair-augmented list.  In every case the Fourier prediction
agrees with Monte Carlo; independent $5\times10^4$-shot checks at $p=0.01$
have maximum deviation $2.6\sigma$ (per-code tables are in the
Supplemental Material~\cite{Supplemental}).
In Fig.~\ref{fig:pmiss}, a point with zero observed misses has $\hat p=0$
and is invisible on the logarithmic axis; with one or a few observed
misses the Wald standard error is comparable to $\hat p$ itself, so the
lower endpoint stays positive but close to zero, producing the long
downward bars at small $p$.

\begin{table}[t]
\caption{Structural metrics for the four test codes.
$n_{\rm dang} = |\{j : M_{{\rm log},j} \neq 0\}|$ is the number of dangerous
columns;
$\Lambda_2^{\rm g}$ is the greedy upper bound on the
unrestricted column-subset size for weight-2 coverage;
$|\mathcal S_{\rm full}|$ is the number of distinct logical displacements in
$\mathcal S_{\rm full}=\mathcal S_0\cup\mathcal S_{\rm pair}$ (baseline,
singles, and all $\binom{q}{2}$ weight-2 pairs).}
\label{tab:cross}
\centering
\footnotesize
\setlength{\tabcolsep}{4pt}
\begin{tabular}{lccccc}
\toprule
Code & $q$ & $k$ & $n_{\rm dang}$ & $\Lambda_2^{\rm g}$ & $|\mathcal S_{\rm full}|$ \\
\midrule
HGP $[\![13,1,3]\!]$   & 7  & 1  & 3  & 0  & 2 \\
HGP $[\![27,4,3]\!]$   & 13 & 4  & 7  & 5  & 15 \\
BB $[\![72,12,6]\!]$   & 42 & 12 & 41 & 37 & 537 \\
BB $[\![144,12,12]\!]$ & 78 & 12 & 67 & 41 & 635 \\
\bottomrule
\end{tabular}
\end{table}

\subsection{Robustness Across Column Orderings}
\label{sec:robust}

Table~\ref{tab:lam2_robust} reports $\Lambda_2^{\rm greedy}$ and the
covered-signature count $|\mathcal{S}_{\rm full}|$ across 30 random column
permutations of the gross code $[\![144,12,12]\!]$.
While the absolute value varies with the ordering (since
reordering changes the RREF and hence $M_{\rm log}$ itself),
$\Lambda_2^{\rm greedy}/q$ stays in $[0.603,0.795]$ across the 30 orderings.
In all 30 orderings, $B_2 = 0$ is achieved (i.e., $\Lambda_2^{\rm greedy}$
always suffices for complete weight-2 coverage).
For the same 30 orderings, the combination-sweep prefix threshold stays near the full free dimension.
The covered-signature count itself ranges over $[892,1384]$ (median 1179;
for the canonical ordering, $|\mathcal{S}_{\rm full}|=635$).

\begin{table}[t]
\caption{$\Lambda_2^{\rm greedy}$ and covered-signature statistics across
30 random column orderings for the gross code; here
$|\mathcal{S}_{\rm full}|=|\mathcal S_0\cup\mathcal S_{\rm pair}|$.}
\label{tab:lam2_robust}
\centering
\begin{tabular}{cccc}
\toprule
Statistic & $\Lambda_2^{\rm greedy}$ & $\Lambda_2^{\rm greedy}/q$ & $|\mathcal{S}_{\rm full}|$ \\
\midrule
Minimum  & 47 & 0.603 & 892 \\
Median   & 56.5 & 0.724 & 1179 \\
Maximum  & 62 & 0.795 & 1384 \\
\bottomrule
\end{tabular}
\end{table}

What persists across orderings is the \emph{coverage-compression
phenomenon}: a small signature-selected column subset suffices for complete
weight-2 coverage, while the standard prefix threshold stays near the full
free dimension.  What is \emph{not} preserved is the identity of the
logical signatures, nor a cross-ordering validation of decoding-error
improvements: changing the column order changes the RREF and the
free-variable representation, so the covered sets and representatives of
different orderings are different objects.

\subsection{Main Comparison: Coverage and Selection on the Gross Code}
\label{sec:main}

We compare four decoders on
the gross (bivariate bicycle) code $[\![144,12,12]\!]$~\cite{Bravyi2024,TourGross2025}
with i.i.d.\ bit-flip noise at physical error rate $p = 0.01$,
using a fixed, error-independent reliability ordering.
All experiments use $N = 5000$ Monte Carlo shots with identical error samples
(seed 7777), enabling paired McNemar tests between decoders.

All four decoders share the same RREF and cost rule, differing only in their
lists: A is $\mathcal F_{\rm CS}(41)$; F is the coverage-minimal construction
of Proposition~\ref{prop:covmin}; B uses all pairs in the 41-column
$\Lambda_2^{\rm greedy}$ set; C is the full $\mathcal F_{\rm CS}(q)$.
Since A and F differ in both coverage and representatives,
A$\to$F is not a pure causal isolation of coverage;
B and~C provide the controlled comparison:
$\mathcal{F}_{\rm B} \subset \mathcal{F}_{\rm C}$,
$\mathcal{S}_{\rm B} = \mathcal{S}_{\rm C}$, yet $P_L$ differs threefold.
B thus replaces F's representative set rather than adding onto it.
Detailed list definitions are in the Supplemental Material~\cite{Supplemental}.

\begin{table}[t]
\caption{Gross-code results at $p=0.01$ on the same 5000 errors.
$|\mathcal S|$ is the number of covered logical displacements and $B_2$ the
number of uncovered weight-2 free patterns.  McNemar values are two-sided
exact tests for consecutive rows.}
\label{tab:main}
\centering
\footnotesize
\setlength{\tabcolsep}{4pt}
\begin{tabular}{lccccc}
\toprule
Decoder & $N_{\rm cand}$ & $|\mathcal{S}|$ & $B_2$ & $P_L$ & McNemar $p$ \\
\midrule
A (CS, $\lambda{=}41$) & 899 & 241 & 823 & 13.20\% & --- \\
F (cov.-min.) & 672 & 635 & 0 & 12.42\% & 0.013 \\
B (sig.-aware) & 899 & 635 & 0 & 11.30\% & $1.2\times10^{-9}$ \\
C (full, $\lambda{=}q$) & 3082 & 635 & 0 & 3.78\% & $5.9\times10^{-110}$ \\
\bottomrule
\end{tabular}
\end{table}

Table~\ref{tab:main} separates the effects of logical-class coverage and
representative selection through controlled pairwise comparisons.
F uses 227 fewer candidates than A, reduces $P_L$ from
$13.20\%$ to $12.42\%$, and eliminates all weight-2 coverage misses.
Although F and B cover the same logical set, replacing the 672-candidate F
list with the 899-candidate B list further reduces $P_L$ to $11.30\%$.
B and C share the same 635 signatures yet differ
threefold in $P_L$ ($11.30\%$ vs.~$3.78\%$).

The event decomposition is exact sample by sample.
B and C share 93 coverage-miss shots; Table~\ref{tab:decomp} splits the
logical error rates into coverage-miss and selection terms.
The Fourier prediction for the same fixed covered set is
$P_{\rm miss}=1.637\%$ at $p=0.01$.

\begin{table}[t]
\caption{Per-shot decomposition of logical failure for decoders B and C
on the 5000 common samples at $p=0.01$.  The two lists have identical
coverage, so the coverage-miss counts coincide.}
\label{tab:decomp}
\centering
\begin{tabular}{lccc}
\toprule
Decoder & $P_{\rm miss}$ & $P_{\rm sel}$ & $P_L$ \\
\midrule
B & 1.86\% & 9.44\% & 11.30\% \\
C & 1.86\% & 1.92\% & 3.78\% \\
\bottomrule
\end{tabular}
\end{table}

In the paired comparison, 378 samples flip from failure with B to success
with C and 2 flip in the reverse direction (exact McNemar
$p=5.88\times10^{-110}$).  Of the 378, 351 are strict rescues, with a
decision margin strictly negative for B and strictly positive for C; the
remaining 27 involve ties, so the count 378 should not be read as
``rescued by margin improvement'' without this qualification.

The margin diagnostics are computed only on the 3720 samples with nonzero
syndrome for which the class energies entering $\Delta$ are finite for
both B and C.  The excluded samples are the 1187 zero-syndrome shots, for
which the margin script performs no candidate enumeration and evaluates no
class energies, and the 93 coverage misses with $E_{y_{\rm true}}=\infty$.
This exclusion is a convention of the margin diagnostics only: zero-syndrome
shots are otherwise processed through the same decoding and logical-error
criterion as all other shots.  On the shots where the energies of the
compared classes are finite, the margin change decomposes as
\begin{equation}
\Delta_C-\Delta_B=\delta_{\rm true}-\delta_{\rm wrong},
\end{equation}
where $\delta_{\rm true}=E_{y_{\rm true}}^B-E_{y_{\rm true}}^C$ is the
energy drop of the true class and
$\delta_{\rm wrong}=\min_{y\ne y_{\rm true}}E_y^B-\min_{y\ne y_{\rm true}}E_y^C$
that of the best competing class; indeed
$\delta_{\rm true}>\delta_{\rm wrong}$ on all 351 rescues.  Adding
representatives can lower the energies of both the true and the competing
classes; whether the decision improves depends on the relative magnitude
of the two drops.

\subsection{Representative-Selection Strategies}

As a test of representative-selection strategies, two further
899-candidate lists keep B's coverage.  Decoder E diversifies within
classes: in addition to one pair per covered signature it selects extra
pairs that maximize the $G_F$ Hamming distance from the anchor correction
within each collision group.  Its error rate $P_L=11.26\%$ is not
statistically significantly different from B's $11.30\%$ (McNemar
$p=0.91$).  Decoder E2 trains the allocation: it covers the signatures as
F does and assigns the remaining 227 pairs to the displacements with the
highest selection-failure counts on a training set of 2000 shots
(seed 12345) disjoint from the test set.  Phase~1 fails on 224 of the
2000 training shots, of which 194 are selection failures within the
covered set (the other 30 are coverage misses under the phase-1 list).
Displacements are ranked by (selection-failure count, first failure
shot), and the 227-pair budget is distributed in proportion to these
counts, each displacement drawing its lowest-weight pairs from its pool
(phase-1 pair excluded).  On an independent 5000-shot test set,
$P_L=11.68\%$, not statistically significantly different from B
($p=0.081$).  The simple allocation
strategies tested here thus do not yet resolve the
representative-selection problem; full details (training sizes, scoring,
quotas, tie rules) are in the Supplemental Material~\cite{Supplemental}.

\section{Discussion}
\label{sec:discussion}

These results identify two separate limitations of a finite
OSD list: the list may omit the true logical class, or it may
contain that class but lose the cost comparison to a competing one---lower
cost, or equal cost with the tie resolved against the true class by the
fixed list order.  Identical coverage does not guarantee identical logical
error rate, as shown by the factor-of-three difference between
decoders~B and~C.

For the gross code, $\lambda_2^*=q$ explains why the combination-sweep list
requires a large pair prefix for complete weight-2 coverage, while $\Lambda_2$
exposes a smaller signature-aware alternative.
Theorem~\ref{thm:2} is exact for fixed, error-independent column orderings,
while Theorem~\ref{thm:3} holds for every realization of any candidate-based
decoder, including data-dependent orderings.  Extending the iid averaging
formulas to BP-informed orderings requires analyzing the induced distribution
on free variables.
The exact minimum of $\Lambda_2$ is a combinatorial optimization problem.
Candidate count alone therefore does not characterize an OSD list.

As a control reported in the Supplemental Material~\cite{Supplemental},
the $\lambda=41$ decoder is repeated with the same engine and candidate
specification, replacing only the fixed column ordering by a BP-informed
reliability ordering (columns ranked by increasing BP posterior LLR; the
decoder always proceeds to OSD, so BP convergence never short-circuits
list selection).  On the same 5000 samples at $p=0.01$ the logical error
rate changes from $13.20\%$ to $0.34\%$.  This ordering sensitivity is
expected.

\emph{Computational cost.}  The tools introduced here carry distinct
costs.  Constructing $M_{\rm log}$ requires one Gaussian elimination of
$H_Z$ and is a one-time offline cost.  Evaluating the exact
logical-signature distribution of Theorem~\ref{thm:2} requires, per
displacement $y$, one sum of $2^k$ terms as in Eq.~\eqref{eq:fourier};
this is at most 4096 terms for the largest code studied here ($k=12$).
Computing the full distribution over all $2^k$ displacements by direct
per-signature summation costs $2^{2k}$ operations, each displacement
requiring a fresh $2^k$-term sum as in Eq.~\eqref{eq:fourier}; the fast
Walsh--Hadamard transform instead costs $k\,2^k$ operations once the
$2^k$ Fourier coefficients entering the sums have been prepared, so the
transform cost does not include that preparation.  Exact evaluation
should therefore not be read as efficient for arbitrarily large $k$.  Beyond the scales studied
here, the small-$p$ expansion of Eq.~\eqref{eq:pmissasym} or Monte Carlo
estimation applies.  Generating the signatures of all $\binom{q}{2}$ pairs
costs $O(q^2k)$ binary operations, one signature computation per pair; the
subsequent greedy column selection, gain evaluation, and coverage updates
are additional and are not characterized by this bound.  Online decoding
evaluates one physical cost per candidate, a total of order
$N_{\rm cand}\,n$ operations once the candidate corrections $G_F f$ have
been precomputed offline; solving for the syndrome baseline $c_0(s)$ is a
separate cost not included here.

These results provide concrete guidance for finite OSD list design. The signature-based constructions reduce candidate count while preserving a target coverage set, and the failure decomposition identifies whether errors arise mainly from missing classes or from selection among covered classes. The B–C comparison shows that compression must also be assessed for its effect on within-class representation, even when coverage is unchanged. The optimal selection of representatives to minimize $P_{\mathrm{sel}}$ at fixed logical-class coverage and candidate budget remains an open problem.

\begin{acknowledgments}
This work is supported by the National Natural Science Foundation special project of China (Grant No. 12341103) and the National Natural Science Foundation of China (Grant No. 62372444), and National Key R\&D Program of China (Grant No. 2023YFA1009403).
\end{acknowledgments}

\end{document}


\title{Supplemental Material for ``Logical-Space Theory of Ordered-Statistics Decoding for Quantum LDPC Codes''}

\author{Yun Shang}
\email{shangyun@amss.ac.cn}
\affiliation{Institute of Mathematics, Academy of Mathematics and Systems Science,
Chinese Academy of Sciences, Beijing 100190, China}
\affiliation{State Key Laboratory of Mathematical Sciences, Academy of Mathematics
and Systems Science, Chinese Academy of Sciences, Beijing 100190, China}

\maketitle

All vectors and matrices below are over $\mathbb F_2$ unless stated
otherwise.  This Supplemental Material fixes the algebraic conventions used
in the main text, reports the BP-informed ordering control and the paired
statistical accounting, specifies the additional heuristic decoders, and
collects the remaining numerical checks underlying the cross-code and
coverage--selection analyses.

\section{CSS conventions and free-variable coordinates}

We consider $X$-error decoding for a CSS code with check matrices
$H_X\in\mathbb F_2^{m_X\times n}$ and
$H_Z\in\mathbb F_2^{m_Z\times n}$ satisfying $H_XH_Z^T=0$.
An $X$ error $e\in\mathbb F_2^n$ has syndrome $s=H_Ze$.
Syndrome-zero $X$ errors form $\ker H_Z$, whereas logically trivial
syndrome-zero errors form $\operatorname{row}H_X$.  Choose a complete set of
logical-$Z$ tests as the rows of $L_Z\in\mathbb F_2^{k\times n}$, so that
\begin{equation}
\ker\!\left(L_Z|_{\ker H_Z}\right)=\operatorname{row}H_X.
\label{eq:S-quotient}
\end{equation}
Thus $L_Z$ induces an isomorphism
$\ker H_Z/\operatorname{row}H_X\cong\mathbb F_2^k$.

Fix a column ordering independent of the realized error.  Let
$r=\operatorname{rank}H_Z$ and $q=n-r$.  After deleting redundant zero rows,
the nonzero part of the RREF is
\begin{equation}
[I_r\mid A],\qquad A\in\mathbb F_2^{r\times q}.
\label{eq:S-rref}
\end{equation}
For each free coordinate $j$, the corresponding nullspace vector in sorted
coordinates is
\begin{equation}
\widetilde g_j=\begin{bmatrix}Ae_j\\e_j\end{bmatrix}.
\end{equation}
After restoring the physical column order, write these vectors as $g_j$ and
define
\begin{equation}
G_F=[g_0\;g_1\;\cdots\;g_{q-1}]\in\mathbb F_2^{n\times q}.
\label{eq:S-GF}
\end{equation}
The map $G_F:\mathbb F_2^q\to\ker H_Z$ is an isomorphism.

For a syndrome $s$, let $c_0(s)$ be a particular solution of $H_Zc_0=s$.
Every error with syndrome $s$ is uniquely expressible as
\begin{equation}
e=c_0(s)+G_Fe_F.
\label{eq:S-error-decomp}
\end{equation}
For the probabilistic results below we use the elimination baseline obtained by
setting all free coordinates to zero.  With this choice, $e_F$ is exactly the
restriction of the physical error to the fixed free coordinates.

\section{BP-informed ordering control}

As a control for the fixed-order experiments, we repeat the $\lambda=41$
decoder on the gross code with the same decoding engine and candidate
specification, replacing only the fixed column ordering by a BP-informed
reliability ordering (min-sum BP with parallel scheduling and
$\alpha=1.0$, columns ranked by increasing posterior LLR; the decoder
always proceeds to OSD, so BP convergence never short-circuits the list
selection).  Table~\ref{tab:bpctrl} reports logical failures on the same
5000 error samples per physical error rate (seed 7777); the improvement is
significant under the paired two-sided exact McNemar test at all three
rates ($p<10^{-3}$).  
\begin{table*}[t]
\caption{BP-informed ordering control on the gross code (5000 error
samples per physical error rate, seed 7777).  ``Fixed order'' is
decoder~A of the main text ($\lambda=41$, 899 candidates).
``BP-informed'' uses the same engine and candidate specification with
BP-posterior column ordering; ``factor'' is the fixed-order failure count
divided by the BP-informed count.  The rightmost columns give the native
ldpc BP+OSD pipeline with the same list size.}
\label{tab:bpctrl}
\centering
\begin{tabular}{cccccc}
\toprule
$p$ & Fix. & BP-inf. & Fct. & BP+OSD & Fct. \\
\midrule
0.01 & 660 (13.20\%) & 17 (0.34\%) & 38.8 & 18 (0.36\%) & 36.7 \\
0.02 & 1840 (36.80\%) & 166 (3.32\%) & 11.08 & 172 (3.44\%) & 10.7 \\
0.03 & 2818 (56.36\%) & 681 (13.62\%) & 4.14 & 693 (13.86\%) & 4.1 \\
\bottomrule
\end{tabular}
\end{table*}

\section{Paired McNemar accounting and decision margin}

The paired two-sided exact McNemar discordant counts between the main
decoders on the same 5000 gross-code shots at $p=0.01$, generated with random seed 7777, are
($b$: shots where the first decoder fails and the second succeeds;
$c$: the reverse discordant count):

\begin{center}
\begin{tabular}{lccc}
\toprule
Comparison & $b$ & $c$ & exact $p$\\
\midrule
A vs F & 137 & 98 & 0.013\\
F vs B & 72 & 16 & $1.19\times10^{-9}$\\
B vs C & 378 & 2 & $5.88\times10^{-110}$\\
\bottomrule
\end{tabular}
\end{center}

For decoders B and C, exactly 93 of the 5000 common samples are coverage
misses; the Fourier prediction for the same fixed covered set is
$P_{\rm miss}=1.637\%$ at $p=0.01$.

The margin diagnostics are computed only on the 3720 samples with nonzero
syndrome for which the class energies entering $\Delta$ are finite for
both B and C.  The excluded samples comprise the 1187 zero-syndrome shots,
for which the margin script performs no candidate enumeration and
evaluates no class energies (a convention of the diagnostics, not a
statement that $\Delta$ fails to be finite), and the 93 coverage misses
with $E_{y_{\rm true}}=\infty$. These zero-syndrome samples are included in the logical-error statistics; all correspond to the zero error in this dataset. B has 446 strict negative-margin failures and 26
additional selection failures lost by tie-breaking at $\Delta=0$; C has 93
strict negative-margin failures and 3 tie failures.  There are 351 strict
rescues with $\Delta_B<0$ and $\Delta_C>0$.  On those shots,
\begin{align}
\delta_{\rm true}&=E_{y_{\rm true}}^B-E_{y_{\rm true}}^C,\\
\delta_{\rm wrong}&=\min_{y\ne y_{\rm true}}E_y^B-
\min_{y\ne y_{\rm true}}E_y^C,
\end{align}
with measured means $82.24$ and $7.19$, respectively, and
$\delta_{\rm true}>\delta_{\rm wrong}$ in all 351 rescues.  The class
energies are evaluated with integer Hamming weights of the corrections
(the per-bit cost $\log(1/p_i)$ is uniform across positions at the
common rate $p=0.01$, so ranking by cost is equivalent to ranking by
physical Hamming weight); the margin differences are quoted in
$\log(1/p)$ cost units by multiplying the integer energy differences
by the uniform factor $\log 100\simeq4.605$, one unit per
$\log(1/p)$-weighted bit flip.  The measured means
correspond to $17.9$ and $1.6$ physical Hamming-weight units,
respectively.

\section{Additional Heuristic Decoders}

\subsection{Decoder E: Coverage + Representative Diversity}

Decoder~E uses the full $\binom{78}{2} = 3003$ weight-2 pairs to construct an
$N_{\rm cand} = 899$ candidate set via a two-phase strategy:
(i) one minimum-intrinsic-weight pair per collision group
(635 pairs, one for each distinct weight-2 displacement; the 593 groups in
$R = \mathcal S_{\rm pair} \setminus \mathcal S_0$, where $\mathcal S_{\rm pair}$
is the set of all pair displacements, supply the coverage beyond
$\mathcal{S}_0 = \{0\} \cup \{\text{single-column sigs}\}$,
while the remaining 42 groups have displacements already realized by
single-column patterns);
(ii) the remaining 185 pairs are selected to maximize $G_F$ Hamming distance
from the anchor correction within each collision group
(groups of pairs sharing the same logical displacement), allowing at most
two additional pairs per group and filling any residual budget with the
highest-scoring remaining pairs (score = anchor distance minus $0.1$ times
intrinsic weight, sorted descending with ties kept in enumeration order).
The resulting logical error rate is $P_L = 11.26\%$,
which is not statistically significantly different from decoder~B
($P_L = 11.30\%$, McNemar $p = 0.91$ on the same 5000 shots).
Diversity heuristics operating on
$G_F$ distance do not produce a statistically resolved improvement over B;
we draw no stronger mechanistic conclusion from decoder~E.

\subsection{Decoder E2: Energy-Optimization Training}

Decoder~E2 uses a training-based allocation.  Phase~1 is identical to
decoder~F: one minimum-intrinsic-weight pair per displacement in
$R = \mathcal S_{\rm pair} \setminus \mathcal S_0$ (593 pairs), with ties
broken by the lexicographic order $(w,i,j)$ of (intrinsic weight, first
column, second column), listed in ascending order of the displacement $y$
interpreted as a $k$-bit integer.  Phase~2 allocates the remaining 227
pairs of the 899-candidate budget to displacements by their phase-1
failure counts on a held-out training set of 2000 shots at $p=0.01$
(seed 12345), disjoint from the 5000-shot test set (seed 7777).
Training labels are the true displacements
$y_{\rm true}=M_{\rm log}e_F$: phase~1 fails on 224 of the 2000 training
shots, of which 194 are selection failures (their displacement lies in
the phase-1 covered set $\mathcal S_{\rm full}$) and 30 are coverage
misses under the phase-1 list.

The allocation order sorts displacements by (selection-failure count,
first failing training shot), descending in count.  For each
displacement the pool of available pairs excludes the phase-1 pair and is
sorted by $(w,i,j)$.  A first pass assigns
$\max(1,\lfloor 227\,N_y/N_{\rm tot}\rfloor)$ pairs per displacement,
capped by the pool size and the remaining budget; a second pass fills
any residual budget in the same displacement order.  The final candidate
order is: the zero pattern, the 78 single-column patterns in free-column
order, the 593 phase-1 pairs, and the phase-2 pairs in allocation order.
Minimum-cost ties are broken by the earliest candidate in this list
order.

The resulting logical error rate on an independent 5000-shot test set is
$P_L = 11.68\%$, not statistically significantly different from decoder~B
($P_L = 11.30\%$, McNemar $p = 0.081$ on the same 5000 shots).
Adding a representative can lower the energy
of either the true or a competing logical class; this allocation rule
does not resolve the representative-selection problem.

\subsection{Decoder F: Coverage-Minimal Baseline}

With the baseline and all single-free-coordinate candidates fixed, decoder~F
uses the minimum possible number of additional pair representatives needed to
realize every signature in $\mathcal{S}_{\rm full}$ and hence achieve $B_2=0$,
attaining the coverage-minimal completion bound of the main text:
$\mathcal{F}_0 \cup$ all weight-1 $\cup$ one minimum-intrinsic-weight
weight-2 pair for each displacement in
$\mathcal{S}_{\rm full} \setminus \mathcal{S}_0$.
The \emph{intrinsic weight} of a weight-2 pair $(i,j)$ is the Hamming weight
$|G_F(e_i+e_j)|$ of the pair's own physical realization (its contribution to
the correction, independent of the syndrome); a minimum-intrinsic-weight pair
for a displacement $y$ minimizes this weight, with ties broken by the
lexicographic order of $(i,j)$.
The candidate list is ordered deterministically: the zero pattern, the 78
single-coordinate patterns in ascending column order, then the 593 pairs in
ascending order of their displacement $y$ interpreted as a $k$-bit integer;
minimum-cost ties in decoding are broken by the earliest candidate in this
list order.
With $N_{\rm cand}=672$, decoder~F achieves
$P_L=12.42\%$, compared with $13.20\%$ for the
899-candidate combination-sweep decoder~A. Thus a coverage-aware list
can outperform the combination-sweep prefix list with fewer candidates.
Because A and F differ in both their covered signatures and
their physical representatives, this comparison is not a pure
causal isolation of coverage.

\begin{table}[t]
\caption{Summary of all decoders tested.
$N_{\rm cand}$: candidate count;
$B_2$: uncovered weight-2 free-variable patterns;
$|\mathcal{S}|$: covered logical displacements.}
\label{tab:all_decoders}
\centering
\footnotesize
\setlength{\tabcolsep}{4pt}
\begin{tabular}{lccccc}
\toprule
Decoder & $N_{\rm cand}$ & $B_2$ & $|\mathcal{S}|$ & $P_L$ \\
\midrule
A (CS, $\lambda{=}41$)  & 899  & 823 & 241 & 13.20\% \\
F (coverage-minimal)     & 672  & 0   & 635 & 12.42\% \\
B (sig.-aware)           & 899  & 0   & 635 & 11.30\% \\
E (coverage + diversity) & 899  & 0   & 635 & 11.26\% \\
E2 (energy training)     & 899  & 0   & 635 & 11.68\% \\
C (full, $\lambda{=}q$)  & 3082 & 0   & 635 & 3.78\%  \\
\bottomrule
\end{tabular}
\end{table}

\section{Collision-Multiplicity Correlation Analysis}

For the gross code, the 635 logical displacements covered by the full
$\binom{78}{2} = 3003$ weight-2 pairs have a highly non-uniform
collision structure: the number of pairs per displacement
(multiplicity $m_y$) ranges from 1 to 120, with median 2 (the zero
displacement $y=0$ itself carries 120 pairs).  The value 4.85
for decoder~C is instead the arithmetic
mean number of \emph{all} candidates per covered signature,
$3082/635\simeq4.85$; it should not be interpreted as the pair-only
multiplicity.  Likewise, decoder~B has $899/635\simeq1.42$ candidates per
covered signature on average.  These arithmetic means describe list density, not decoding performance.

A natural question is whether the $\Lambda_2^{\rm greedy}$ clique
preserves the collision structure of the full pair set---i.e., whether
the displacements that have many representatives under full pairs
also have many under the clique.

The measured multiplicities $m_y^{\rm B}$ and $m_y^{\rm full}$ have Pearson
correlation $r = 0.254$, indicating that the clique
\emph{does not} preserve the collision structure.
We therefore do not interpret the performance of decoder~B as evidence that
the greedy clique preserves a natural collision prior.  The F--B comparison shows that two lists with the same
covered-signature set can have different selection performance
when their physical representatives differ. It does not isolate
the effect of candidate number, because $\mathcal F_F$ need not
be a subset of $\mathcal F_B$. The nested B--C comparison,
$\mathcal F_B\subset\mathcal F_C$ with
$\mathcal S_B=\mathcal S_C$, provides the controlled evidence
that additional within-class representatives can improve
selection.  Which representatives are optimal is a separate candidate-design
problem.

\section{Additional Cross-Code Metrics}

For all four test codes, $\operatorname{rank}(M_{\rm log}) = k$ is
verified, confirming Theorem~1 of the main text.  For the full
pair-augmented lists, $B_w$ and $t_{\rm cov}$ are computed against
$\mathcal S_{\rm full}=\mathcal S_0\cup\mathcal S_{\rm pair}$ (baseline,
singles, and all pair displacements); at $p=0.01$ the corresponding exact
Fourier predictions ($P^{\rm th}$) and independent coverage Monte Carlo
checks ($P^{\rm MC}$, $5\times10^4$ samples per code) are

\begin{center}
\footnotesize
\setlength{\tabcolsep}{4pt}
\begin{tabular}{lcccc}
\toprule
Code & $B_3$ & $t_{\rm cov}$ & $P^{\rm th}$ & $P^{\rm MC}$ \\
\midrule
HGP $[\![13,1,3]\!]$ & 0 & $\infty$ & 0 & 0\\
HGP $[\![27,4,3]\!]$ & 7 & 3 & $7\times10^{-6}$ & $2\times10^{-5}$\\
BB $[\![72,12,6]\!]$ & 6705 & 3 & 0.00514 & 0.00494(31)\\
BB $[\![144,12,12]\!]$ & 25278 & 3 & 0.01637 & 0.01790(59)\\
\bottomrule
\end{tabular}
\end{center}

In $P^{\rm MC}$ the parenthetical is the $\pm1$ Wald standard error
$\sqrt{\hat p(1-\hat p)/N}$ of the binomial proportion with
$N=5\times10^4$ (not a confidence interval).
For BB72 and the gross code the deviations are approximately $0.6\sigma$ and
$2.6\sigma$, respectively.  The exact theory may be evaluated arbitrarily far
into the low-$p$ regime; Monte Carlo comparisons are interpreted only where
coverage misses are statistically resolved by the finite sample.

\section{Coverage-deficit spectrum}

The $B_w$ spectrum for the gross code
under the canonical reliability-sorted ordering, computed for the
full pair-augmented list
($\mathcal{S}_{\rm full}=\mathcal{S}_0\cup\mathcal{S}_{\rm pair}$):

\medskip
\begin{center}
\refstepcounter{table}\textbf{Table \thetable.}
$B_w$ spectrum for the gross code under canonical ordering.
$B_w = |\{x \in \mathbb{F}_2^{78} : |x| = w, M_{\rm log} x \notin \mathcal{S}_{\rm full}\}|$.
$B_0 = B_1 = B_2 = 0$ by construction of $\mathcal{S}_{\rm full}$.
$B_3$ is computed exactly.
\par\medskip
\begin{tabular}{ccc}
\toprule
$w$ & $B_w$ & $\binom{78}{w}$ \\
\midrule
0 & 0 & 1 \\
1 & 0 & 78 \\
2 & 0 & 3003 \\
3 & $25\,278$ & 76\,076 \\
\bottomrule
\end{tabular}
\end{center}
\medskip

$t_{\rm cov} = 3$ confirms that weight-2 coverage is complete, and the
first uncovered contributions enter at order $p^3$.
The small-$p$ asymptotic
$P_{\rm miss} \approx 25278\,p^3$ is accurate for
$p \lesssim 10^{-3}$.  At $p=0.01$, the bare asymptotic value
$25278p^3\simeq0.0253$ is no longer a quantitative approximation: the exact
weight-three contribution includes the factor $(1-p)^{75}$, while weights
$w\ge4$ also contribute.  Those higher-weight terms need not be enumerated
individually: their combined effect is contained in the exact Fourier value
$P_{\rm miss}=0.01637$.  At $p=0.01$ the full formula, not the asymptotic
one, is required.